%Paper: hep-th/9309066
%From: SHIBAJI@ITSICTP.BITNET
%Date: Fri, 10 Sep 93 18:01 N

\magnification=1200
\font\titlea=cmb10 scaled\magstep1
\font\titleb=cmb10 scaled\magstep2
\rightline{IC/93/307}
\rightline{hepth@xxx/9309066}
\baselineskip=18pt
\vskip .5cm
\centerline{\titleb Topological Conformal Algebra in ${\bf 2d}$ Gravity}
\centerline{\titleb Coupled to Minimal Matter}
\vskip 1.0cm
\baselineskip=14pt
\centerline{Sudhakar Panda}
\smallskip
\centerline{\it The Mehta Research Institute of Mathematics}
\centerline{\it and Mathematical Physics}
\centerline{\it Allahabad 211 002, India}
\bigskip
\centerline{and}
\bigskip
\centerline{Shibaji Roy}
\smallskip
\centerline{\it International Centre for Theoretical Physics}
\centerline{\it Trieste, Italy}
\vskip 1.0cm
\baselineskip=12pt
\centerline{\titlea ABSTRACT}
\bigskip
{\parindent=3pc
\narrower{ An infinite number of topological conformal
algebras with varying central
charges are explicitly shown to be present in $2d$ gravity (treated both in
the conformal gauge and in the light-cone gauge) coupled to minimal matter.
The central charges of the underlying $N=2$ theory in two different gauge
choices are generically found to be different. The physical states in these
theories are briefly discussed in the light of the $N=2$ superconformal
symmetry.\smallskip}}
\vskip 2.0cm
\baselineskip=14pt
Despite the much recent efforts to understand the results of the discretized
version of $2d$ gravity coupled to various matter systems (matrix models) in
terms of the continuum approach, many questions remain unresolved [1,2].
One such
question is the origin of a topological structure (which is present in the
matrix models [3,4]) directly in the conventional approach of Liouville-matter
system. Only known field theoretic description of the matrix model formulation
of $2d$ gravity is the $2d$ topological gravity coupled to topological matter
[5].
Although it is well-known that some of the matrix model results can be
reproduced [6] in the continuum approach of Liouville-matter system, yet the
topological structure of the latter were not understood until recently. In
ref.[7], it is shown that almost all string theories,
including the bosonic string,
the superstring and $W$-string theories possess a topological conformal algebra
(TCA). This is certainly an indication of a possible connection between the
topological field theories and the conventional Liouville-matter system.

By suitably modifying the generators in $2d$ gravity coupled to minimal matter
[8] we explicitly show here that there are in fact infinite number of
TCA's with
varying central charges. We have treated $2d$ gravity both in the conformal
gauge [9] and in the light-cone gauge [10]. The central charges
associated with the
underlying $N=2$ theory for the two gauge choices are found not to be the same.
This shows that there might be an ambiguity in the analysis of the physical
states by relying on the $N=2$ symmetry alone. We, however, discuss very
briefly
the physical states in these theories only when $2d$ gravity is treated in the
conformal gauge.

In the conformal gauge, the conformal degree of freedom of the metric is taken
as the Liouville field and the gravity sector is realized by the Liouville
action. The $(p,q)$ minimal models (with gcd $(p,q)$=1) coupled to Liouville
field can be described in terms of the Coulomb gas representation with the
energy-momentum tensors for the matter and the Liouville sector given as,
$$
\eqalign{ T_M(z) &= -{1\over 2} :\partial\phi_M(z)\partial\phi_M(z): + i Q_M
\partial^2\phi_M(z)\cr
T_L(z) &= -{1\over 2} :\partial\phi_L(z) \partial\phi_L(z): + i Q_L \partial^2
\phi_L(z)\cr}\eqno(1)
$$
where $\phi_M$, $\phi_L$ represent matter and Liouville fields respectively.
$2 Q_M$, $2 Q_L$ are the corresponding background charges. The matter sector
is characterized by the Virasoro central charge $ 1-{6(p-q)^2\over pq} =
1-12Q_M^2$. Since, the total central charge of the Liouville-matter system
should be 26, we find
$$
\eqalign{2 Q_M &= \sqrt{{2p\over q}} - \sqrt{{2q\over p}}\cr
2 Q_L &= i \left(\sqrt{{2p\over q}} + \sqrt{{2q\over p}}\right)\cr}\eqno(2)
$$
The BRST current for this system is given as,
$$
J_B(z) = :c(z)\left[ T_M(z) + T_L(z) + {1\over 2} T^{bc}(z)\right]:\eqno(3)
$$
Here $T^{bc}$ is the energy-momentum tensor for the reparametrization ghost
system, consisting of the ghost field $c(z)$ and the antighost field $b(z)$
with conformal weight $-1$ and 2 respectively and is given by,
$$
T^{bc}(z) = - 2 :b(z)\partial c(z): - :\partial b(z) c(z):\eqno(4)
$$
It has been observed before that the generators $T(z) \equiv T_L(z) + T_M(z)
+ T^{bc}(z)$ ; $ G^+(z) \equiv J_B(z)$ ; $G^-(z) \equiv b(z)$ and $J(z)
\equiv :c(z) b(z):$ satisfy an almost TCA, but the algebra does not close
and produce two new fields $c(z)$ and $c\partial c(z)$ [11].

It is, however, possible to modify the generators $G^+(z)$ and $J(z)$ by
adding total derivative terms [7] (it does not affect the BRST charge) in such
a way that the modified generators would form a closed TCA. The most general
modifications consistent with the conformal weight and ghost charge are given
as
$$
\eqalign{ G^+(z) &= J_B(z) + a_1 \partial(c\partial\phi_L)(z) + a_2 \partial
(c\partial \phi_M)(z) + a_3 \partial^2 c(z)\cr
J(z) &= :c(z) b(z): + a_4 \partial \phi_L (z) + a_5 \partial\phi_M(z)\cr}
\eqno(5)
$$
where $a_i~~(i=1,2,3,4,5)$ are arbitrary parameters. It is now easy to check
the the new generators form a TCA [11]
$$
\eqalign{ T(z) T(w) & \sim {2 T(w)\over (z-w)^2}
+ {\partial T(w) \over (z-w)}\cr
T(z) G^{\pm}(w) & \sim {{1\over 2} (3 \mp 1) G^{\pm}(w) \over (z-w)^2} +
{\partial G^{\pm}(w) \over (z-w)}\cr
T(z) J(w) & \sim {-{1\over 3} c \over (z-w)^3} + {J(w) \over (z-w)^2} +
{\partial J(w) \over (z-w)}\cr
J(z) G^{\pm}(w) & \sim \pm {G^{\pm}(w) \over (z-w)}; \qquad J(z) J(w) \sim
{{1\over 3} c\over (z-w)^2}\cr
G^{+}(z) G^{-}(w) &\sim {{1\over 3} c\over (z-w)^3} + {J(w)\over (z-w)^2}
+ {T(w) \over (z-w)}\cr
G^{\pm}(z) G^{\pm}(w) & \sim 0\cr}\eqno(6)
$$
provided $a_i$'s satisfy
$$
\eqalign{& a_1 + a_4 = 0\cr
& a_2 + a_5 = 0\cr
& a_1^2 + a_2^2 + 2 a_3 -1 = 0\cr
& 2iQ_M a_2 + 2iQ_L a_1 - 2a_3 + 3 = 0\cr}\eqno(7)
$$
The central charge of the associated $N=2$ theory is $c = 6 a_3$. Because
there are three unknown parameters namely, $a_1$, $a_2$ and $a_3$ with two
independent equations governing them in (7), there are infinite number of
solutions for $a_1$ and $a_2$. Consequently, there are infinite number of TCAs
with central charges $6a_3$ present in $2d$ gravity coupled to minimal matter.
In ref.[7] a particular solution of Eq.(7) i.e. $a_2 = 0$ were chosen.
In this case, we have $a_1 = \sqrt{{2p \over q}}$ and $c = 6a_3 = 3(1-{2p\over
q})$ and consequently, there are two $N=2$ superconformal algebra for fixed
values of $p,q$ (and interchanging $p$ and $q$ everywhere in the above).
However,
it has been pointed out in ref.[12] that there is a problem in choosing the
current $\partial\phi_L$ to modify the generators $G^+$ and $J$ when the
cosmological constant is non-zero. This situation will correspond to choosing
$a_1 = 0$. Therefore, we have $a_2 = i\sqrt{{2p\over q}}$ and $c = 3(1+ {2p
\over q})$ and we will be again left with only two TCAs.

In the light-cone gauge, the metric degrees of freedom are fixed by $h_{+-}
= h_{-+} = {1\over 2}$ and $h_{--} = 0$. As shown in ref.[10], the non-zero
components of the metric admits a decomposition in terms of the three
generators of the non-compact group $SL(2,R)$ satisfying the current algebra
$$
j^a(z) j^b(w) \sim {f^{ab}_{~~c} j^c(w)\over (z-w)} + {{k\over 2} \eta^{ab}
\over (z-w)^2}\eqno(8)
$$
where $a,b = 0,\pm$ are $SL(2,R)$ indices, $k$ is the level of the current
algebra, the non-zero components of the killing metric and the structure
constants are given as $\eta^{+-} = \eta^{-+} = -2\eta^{00} = 2$; $
f^{0+}_{~~+} = -f^{0-}_{~~-} = -{1\over 2}$; $f^{+-}_{~~0} = -1$. The residual
gauge invariance is generated by the current $j^+(z)$ and the energy-momentum
tensor $T_G(z)$. The latter is given by the modified Sugawara form [10]
$$
T_G(z) = {1\over k-2} :\eta_{ab} j^a(z) j^b(z): -\partial j^0(z)\eqno(9)
$$
and the associated Virasoro central charge is ${3k\over k-2} + 6k$. With
respect to this energy-momentum tensor the currents $j^+$, $j^0$ and $j^-$
have conformal weights 0, 1 and 2 respectively. The total energy-momentum
tensor when minimal matter is coupled to light-cone gauge gravity is given as,
$$
T(z) = T_G(z) + T_M(z) + T^{bc}(z) + :\partial\zeta \epsilon(z):\eqno(10)
$$
where the extra fermionic ghost system $(\zeta, \epsilon)$
having conformal weights
(0,1) is the consequence of the symmetry associated with the generator $j^+$.
The Virasoro central charge for this ghost system is $-2$. The expression for
the BRST current has the form [13]
$$
J_B(z) = :c(z)\left[ T_G(z) + T_M(z) + {1\over 2} T^{bc}(z) + T^{\zeta
\epsilon}(z)\right]: +\epsilon(z) j^+(z)\eqno(11)
$$
with $T^{\zeta \epsilon}(z) = :(\partial \zeta) \epsilon(z):$.

As in the conformal gauge, the generators $T(z)$, $G^+(z)\equiv J_B(z)$,
$G^-(z)\equiv b(z)$ and $J(z) \equiv :c(z)b(z): + :\epsilon(z) \zeta(z):$
satisfy an almost TCA. The operator product $J_B(z) J_B(w)$ in this case
produce apart from $c(z)$, $c\partial c(z)$ an extra field $c\epsilon j^0(z)$.
In analogy with the conformal gauge case, we here modify the generators as
follows,
$$
\eqalign{G^+(z) &= J_B(z) + A_1 \partial(c\zeta\epsilon)(z) + A_2 \partial^2
c(z) + A_3 \partial(cj^0)(z) + A_4 \partial(c\partial\phi_M)(z)\cr
J(z) &= :c(z) b(z): + A_5 :\epsilon(z) \zeta(z): + A_6 j^0(z) + A_7 \partial
\phi_M(z)\cr}\eqno(12)
$$
with $J_B$ as given in (11). We find that these new generators form TCA Eq.(6)
provided $A_i$'s obey the following relations,
$$
\eqalign{& A_1 - A_5 = 0\cr
& A_3 + A_6 = 0\cr
& A_4 + A_7 = 0\cr
& A_1 + A_3 - 1 = 0\cr
& A_1 + 2 A_2 + k A_3 - 2 i Q_M A_4 - 3 = 0\cr
&2 A_1^2 + 4 A_1 + 4 A_2 + k A_3(4-A_3) - 2 A_4 (A_4+4iQ_M) - 10 = 0\cr}
\eqno(13)
$$
and the central charge of the associated $N=2$ theory is given by $c=6A_2$.
Again we notice that there are three independent unknown parameters ($A_1$,
$A_2$ and $A_4$), but two relations governing them. One can fix $A_1$ and $A_2$
in terms of $A_4$ and so for different values of $A_4$ we have a TCA with
different central charges. Using the central charge balance equation for the
light-cone gauge gravity coupled to matter system, namely,
$$
{3k\over k-2} + 6k + 1 - {6(p-q)^2 \over pq} - 26 - 2 = 0\eqno(14)
$$
we can obtain $k$ in terms of $p,q$ as $k = {p\over q} + 2 $ or $k = {q\over p}
+ 2$. Substituting this value of $k$ in the particular case when $A_4 = 0$
(this corresponds to the case in ref.[7]) we find that the central charge of
the
$N=2$ theory has values
$$
c = 6\left({p\over q} - {q\over p} + 1\right)\qquad {\rm or}\qquad = 6\eqno(15)
$$
The second solution is a particular case of the first when $p = q = 1$ and
corresponds to $c_M$=1 matter coupled to gravity. Comparing the corresponding
expression in the conformal gauge for $c$ which is $c = 3(1-{2p\over q})$ we
note that the underlying $N=2$ theories are different for two different gauge
choices of the metric. In fact this is true for the generic case also. We,
therefore, conclude that unless we can establish an automorphism under which
the
generators of the $N=2$ algebra in these two gauges have one to one
correspondence and the central charge is the same in both cases it might be
ambiguous to determine the physical states by relying on the $N=2$ symmetry
alone.

In the following we make a few remarks about the physical states in the light
of the underlying $N=2$ symmetry only when the gravity is treated in the
conformal gauge. Physical states in the Liouville-matter system are the states
which are in the kernel of the BRST charge $Q_B = \oint dz J_B(z)$ with
$J_B(z)$ as given in (3) modulo its image. It is well known that the physical
state spectrum in this model consists of apart from the usual ghost number zero
states infinite other states with higher ghost numbers [14]. Using the
state-operator correspondence, it has been found in ref.[15] that the ghost
number zero operators (ghost number $-1$ states) define an interesting ring
structure the so-called ``ground ring". For the general $(p,q)$ model coupled
to gravity they have the form [16]
$$
\eqalign{ x &= \left[bc-\sqrt{{p\over 2q}}(i\partial\phi_M - \partial\phi_L)
\right] e^{i\sqrt{{q\over 2p}}(\phi_M-i\phi_L)}\cr
y &= \left[bc + \sqrt{{q\over 2p}}(i\partial\phi_M + \partial\phi_L)\right]
e^{-i\sqrt{{p\over 2q}}(\phi_M + i\phi_L)}\cr}\eqno(16)
$$
Since all the higher ghost number states fall in the module of the
ground ring [17],
one can consider only the ground ring generators. It has been noted in ref.[7],
when $a_2 =0$ in (7) that the central charge becomes the same as the unitary
minimal $N=2$ theory for $p=1$ and $q=l+2$. In this case one finds that $y$
becomes a chiral primary field [18] satisfying the relation
${1\over 2} q_y = h_y$,
where $q_y$ is the $U(1)$ charge and $h_y$ is the conformal weight of $y$.
But $x$
is not a primary field with respect to the $N=2$ theory. Since the unitary
minimal $N=2$ theory is characterized by the ring relation $y^{l+1} = 0$ [18],
which
is also present in $(1,l+2)$ model coupled to gravity one readily identifies
these models with $M_{1,l+2}$ models coupled to gravity.

In general, when $a_2 \neq 0$, we find that the ground ring generators have
$U(1)$ charges $q_x = \sqrt{{q\over 2p}}(a_1 + i a_2)$ and $q_y = \sqrt
{{p\over 2q}}(a_1-ia_2)$ and they have conformal weights $h_x = {1\over 2}q_x$,
$h_y = {1\over 2}q_y$ respectively. We, however, find that in general $x$ and
$y$ are not primary fields since their OPE with the untwisted energy-momentum
tensor are anomalous. By looking at the anomaly terms which are proportional to
$\sqrt{{p\over 2q}}(a_1-ia_2) - 1$ and $\sqrt{{q\over 2p}}(a_1 + ia_2) - 1$ for
$x$ and $y$, it is clear that it is not possible to make both them primary,
since the parameters $a_1$ and $a_2$ also have to satisfy Eq.(7). Since
for general $a_1$, $a_2$, the underlying $N=2$ theory is non-unitary, we need
more detailed investigation in order to draw any conclusion about the physical
states in this case.
\vskip 1cm
\noindent{\titlea Acknowledgments:}
\vskip .5cm
One of us (S.R.) would like to thank Prof. A. Salam, IAEA and UNESCO at the
International Centre for Theoretical Physics, Trieste, for hospitality and
support.
\vskip 1cm
\noindent{\titlea References:}
\vskip .5cm
\item{1.} P. Ginsparg, in {\it proceedings of Trieste Summer School 1991 and
references therein.}
\item{2.} P. Ginsparg and G. Moore, {\it proceedings of TASI 92},
eds. J. Harvey and J. Polchinski.
\item{3.} E. Witten, {\it Nucl. Phys.} {\bf B340} (1990) 281; {\it Surv. Diff.
Geom.} {\bf 1} (1991) 243.
\item{4.} M. Kontsevich, {\it Max Planck Institute preprint} MPI/91-77.
\item{5.} K. Li, {\it Nucl. Phys. } {\bf B354} (1991) 711, 725.
\item{6.} M. Goulian and M. Li, {\it Phys. Rev. Lett.} {\bf 66} (1991) 2051;
P. DiFrancesco and D. Kutasov, {\it Phys. Lett. } {\bf B261} (1991) 385;
M. Bershadsky and I. Klebanov, {\it Nucl. Phys.} {\bf B360} (1991) 559.
\item{7.} M. Bershadsky, W. Lerche, D. Nemeschansky and N. Warner,
{\it Nucl. Phys.} {\bf B401} (1993) 304.
\item{8.} S. Panda and S. Roy, {\it ICTP preprint} IC/93/81, UG-3/93.
\item{9.} F. David, {\it Mod. Phys. Lett.} {\bf A3} (1988) 1651; J. Distler
and H. Kawai, {\it Nucl. Phys. } {\bf B321} (1989) 509.
\item{10.} A. M. Polyakov, {\it Mod. Phys. Lett.} {\bf A2} (1987) 893; V. G.
Knizhnik, A. M. Polyakov and A. B. Zamolodchikov, {\it Mod. Phys. Lett.}
{\bf A3} (1988) 819.
\item{11.} R. Dijkgraaf, E. Verlinde and H. Verlinde, preprint PUPT-1217,
IASSNS-HEP-90/80.
\item{12.} S. Mukhi and C. Vafa, preprint HUTP-93/A002, TIFR/TH/93-01.
\item{13.} T. Kuramoto, {\it Phys. Lett.} {\bf B233} (1989) 363; Z. Horvath,
L. Palla and P. Vecsernyes, {\it Int. Jour. Mod. Phys.} {\bf A4} (1989) 5261;
K. Itoh, {\it Nucl. Phys.} {\bf B342} (1990) 449.
\item{14.} B. Lian and G. Zuckerman, {\it Phys. Lett.} {\bf B254} (1991) 417.
\item{15.} E. Witten, {\it Nucl. Phys.} {\bf B373} (1992) 187.
\item{16.} D. Kutasov, E. Martinec and N. Seiberg, {\it Phys. Lett.} {\bf B276}
(1992) 437.
\item{17.} S. Govindarajan, T. Jayaraman and V. John, preprint IMSc-92/30.
\item{18.} W. Lerche, C. Vafa and N. P. Warner, {\it Nucl. Phys.} {\bf B324}
(1989) 427.
\end